\documentclass[journal]{IEEEtran}
\usepackage{cite}

\ifCLASSINFOpdf
\else
\fi
\usepackage{amsmath}
\usepackage{url}

\usepackage{amssymb}
\usepackage{mathtools}
\usepackage{bm}
\usepackage{siunitx}
\usepackage{enumitem}

\makeatletter
\newcommand*{\rom}[1]{\expandafter\@slowromancap\romannumeral #1@}
\makeatother

\usepackage{tikz}
\usepackage{pgfplots}
\pgfplotsset{compat=newest}

\newtheorem{corollary}{Corollary}

\newtheorem{lem}{Lemma}
\newtheorem{theorem}{Theorem}

\newtheorem{definition}{Definition}

\newcommand{\eu}{\mathrm{e}}

\newcommand{\E}{\mathbb{E}}

\newcommand{\X}{\mathbf{X}}
\newcommand{\Z}{\mathbf{Z}}
\newcommand{\Y}{\mathbf{Y}}
\newcommand{\y}{\mathbf{y}}
\newcommand{\x}{\boldsymbol{x}}
\newcommand{\I}{\boldsymbol{\mathsf{I}}}
\newcommand{\U}{\mathbf{U}}

\newcommand{\A}{ \boldsymbol{\mathsf{A}}}

\newcommand{\C}{ \boldsymbol{\mathsf{C}}}
\newcommand{\HH}{ \boldsymbol{\mathsf{H}}}
\newcommand{\blambda}{ \boldsymbol{\lambda}}

\newcommand{\diag}{ \boldsymbol{\mathsf{diag}}}

\title{The Vector Poisson Channel:  On  the Linearity of the Conditional Mean Estimator }%
\author{ Alex~Dytso,~\IEEEmembership{Member,~IEEE,} 
Michael~Fau\ss,~\IEEEmembership{Member,~IEEE,}
               and~H. Vincent~Poor,~\IEEEmembership{Life~Fellow,~IEEE}
\thanks{ A.~Dytso, M. Fau\ss, and H.~V.~Poor are with the Department of Electrical Engineering, Princeton University, Princeton, NY 08544, USA. E-mail:  adytso@prinction.edu, mfauss@princeton.edu, poor@princeton.edu.}
\thanks{This work was supported by the U.S. National Science Foundation under Grant CCF-1908308.}
}

\IEEEoverridecommandlockouts      

\begin{document}
\maketitle

\begin{abstract}
This work studies properties of the conditional mean estimator in vector Poisson noise. 
The main emphasis is to  study conditions on  prior distributions that induce linearity of  the conditional mean estimator. The paper consists of two main results.   
 The first result shows that the only distribution that  induces  the linearity of the conditional mean estimator is a product gamma distribution.  Moreover, it is shown that the conditional mean estimator  cannot be linear when the dark current parameter of the Poisson noise is non-zero. 
The second result produces a quantitative refinement of the first result.  Specifically, it is shown that  if  the conditional mean estimator is close to linear in a mean squared error sense, then the prior distribution must be close to a product gamma distribution in terms of their characteristic functions. 
 Finally, the results are compared to their Gaussian counterparts.  
\end{abstract}

\section{Introduction} 
This work considers a problem of estimating a random vector $\X$  from a noisy observation $\Y$    where  $\Y$ given $\X=\boldsymbol{x}$ (denoted by $\Y|\X=\boldsymbol{x}$) follows a \emph{vector Poisson distribution}.  The objective is to characterize conditions under which the  \emph{conditional mean estimator} (i.e., $\E[\X| \Y] $)  is a linear estimator.  Conditional mean estimators are an important class of estimators that are optimal under a large family loss functions, namely Bregman divergences \cite{banerjee2005optimality}.  
For example,  we are interested in characterizing the set of prior distributions on $\X$ that induce  linearity of $\E[\X| \Y] $. Also, we are interested in which linear estimators are realizable from   $\E[\X| \Y] $. That is, given that  $\E[\X| \Y] =\HH \Y+\boldsymbol{c}$,  what values of a matrix $\HH$ and vector  $\boldsymbol{c}$ are permitted?  Finally, we are interested in the question of the stability of linear estimators. In other words, suppose that  $\E[\X| \Y]$ is `close' to a linear function, can we make statements about the distribution of $\X$?

Note that the aforementioned questions have been answered for the Gaussian noise model and are part of standard tools of statistical signal processing. Despite the wide use of the Poisson noise model in statistical science, such questions have not been fully addressed in the vector Poisson case. The aim of this work is to fill this gap. 

The linearity of the conditional expectation is intimately connected with a notation of \emph{conjugate priors}, which is an important element of Bayesian statistics.    In its original definition in \cite[Ch.~3]{raiffa1961applied}, the family of prior distributions  is said to be conjugate if it is closed under sampling --  the prior is said to be closed under sampling when both prior and posterior belong to the same family of distributions.  In other words, the distribution of $\X$  and the distribution of $\X|\Y=\boldsymbol{y}$ are in the same family.   

The structure of the conjugate prior is   highly dependent on the nature of the distribution   of $\Y|\X=\boldsymbol{x}$ (often termed likelihood distribution or noise distribution).   For example, in \cite{diaconis1979conjugate}, authors have made  considerable progress in characterizing conjugate priors for the case when the likelihood distribution belongs to the \emph{exponential family}.  In particular, in \cite{diaconis1979conjugate}, it has been shown that a subset  of  the exponential family,  characterized by certain regularity conditions,  has a corresponding set of conjugate priors. Moreover, this set of conjugate priors is completely characterized by the linearity of the posterior expectation: 
\begin{align}
\E[\X| \Y] = \HH   \Y +{\bf b}, 
\end{align} 
where $\HH=a \I$ for  some constant  $a$ and ${\bf b}$ is some constant vector.     

We note that the case when $\HH$ is  a general matrix was not considered in \cite{diaconis1979conjugate}. 
Moreover,  even the case  when $\Y|\X=\boldsymbol{x}$ follows a Poisson distribution is not covered by  the regularity conditions found in \cite{diaconis1979conjugate}.  However, it was  shown earlier in \cite{johnson1957uniqueness} that the conjugate prior for the scalar Poisson distribution is a gamma distribution, and that the linearity  of the posterior expectation holds and is a characterizing property.  The proof in \cite{johnson1957uniqueness} was generalized in \cite{chou2001characterization} to include several families of discrete distributions not covered by the regularity conditions of \cite{diaconis1979conjugate}.     This work considers an arbitrary matrix  $\HH$  and characterizes the sufficient and necessary conditions for the existence of the conjugate prior.

The literature on the Poisson distribution is considerable, and the interested reader is referred to   \cite{CompresseSensingPoisson,wang2015signal} and \cite{wang2013designed} for applications of the Poisson model in compressed sensing;  \cite{verdu1999poisson} and  \cite{lapidoth2009capacity} for a summary of communication theoretic applications;   \cite{guo2008mutualPoisson,atar2012mutual} and  \cite{wang2014bregman}  for applications  in information theory; and   \cite{grandell1997mixed}, \cite{StochApproxPoissonWang} and \cite{dytso2019estimation} for applications of the Poisson distributions in signal processing and other fields.

The paper is organized as follows.  Section~\ref{sec:Model} presents the Poisson noise model.  Section~\ref{sec:MainResults} presents and discusses our main results, which are described in Theorem~\ref{thm:Main1} and Theorem~\ref{thm:QuantitativeRefinement}.       Section~\ref{sec:thm:Main1} and Section~\ref{proof:thm:QuantitativeRefinement} are dedicated to the proofs of Theorem~\ref{thm:Main1} and Theorem~\ref{thm:QuantitativeRefinement}, respectively. Finally, Section~\ref{sec:Applications} concludes the paper and discusses implications of our results by reflecting on the following:   a practically relevant parametrization of a Poisson noise model, which, for example, explicitly incorporates the dark current parameter; and   Gaussian noise counterparts of our results.

\paragraph*{Notation}    Throughout the paper we adopt the following notation.   $\mathbb{R}^n$ denotes the space of all $n$-dimensional vectors, $\mathbb{R}^k_{+}$ the space of all $n$-dimensional vectors with non-negative components,  and $\mathbb{Z}^n_{+} $ the $n$-dimension non-negative integer lattice.   Vectors are denoted by bold lowercase letters, random vectors by bold uppercase letters, and matrices by bold uppercase sans serif letters (e.g., $\boldsymbol{x}, \X,  \boldsymbol{\mathsf{X}}$).  All vectors are are assumed to be column vectors.   For $\boldsymbol{x} \in \mathbb{R}^n$, $\diag(\boldsymbol{x}) \in \mathbb{R}^{n \times n}$ denotes the diagonal matrix with the main diagonal given by $\boldsymbol{x}$.  The vector  with one at position $i$ and zero otherwise is denote by $\boldsymbol{1}_{i}$. In this paper, the gamma distribution has a probability density function (pdf) given by 
\begin{align}
f(x)= \frac{\alpha^\theta}{ \Gamma(\theta)  }  x^{\theta-1} \eu^{-\alpha x},\, x \ge 0, \label{eq:pdfGamma}
\end{align}
 where $\theta>0$ is the shape parameter and $\alpha>0$ is the rate parameter. 
 We denote the distribution with the pdf in \eqref{eq:pdfGamma} by $\mathsf{Gam}(\alpha, \theta)$.

\section{Poisson Noise Model}  
\label{sec:Model}
Let  $\Y \in \mathbb{Z}^k_{+}$ and $\X \in \mathbb{R}^n$. We say that $\Y$ is an output of a system with Poisson noise, if  $\Y| \X=\x$ follows a Poisson distribution, that  is,
\begin{align}
P_{\Y|\X}(\y|\x)= \prod_{i=1}^k  P_{Y_i|\X}(y_i|\x) \label{eq:PoissonChannel}
\end{align}
where 
\begin{align}
 P_{Y_i|\X}(y_i|\x)= \frac{1}{y_i!} ( [\A\x]_i  +\lambda_i   )^{ y_i } \eu^{-( [\A\x]_i +\lambda_i )}, \label{eq:SclarNoise}
\end{align}
$\A \in \mathbb{R}^{k  \times n}$ and $\blambda=[\lambda_1, \ldots, \lambda_k]^T \in  \mathbb{R}^k_{+} $. 
 In \eqref{eq:SclarNoise} we use the convention that $0^0=1$. 
 
  Using the terminology of laser communications, we refer to $\A$ as the \emph{intensity matrix} and $\blambda$ as the \emph{dark current} vector.  
Moreover, we assume that the matrix $\A$ must satisfy the following non-negativity preserving constraint: 
\begin{align}
\A \x \in \mathbb{R}^k_{+} , \,  \forall  \x \in  \mathbb{R}^n_{+} . 
\end{align}

The random transformation of the input random variable $\X$ to an output random variable $\Y$  by  the channel in \eqref{eq:PoissonChannel} is denoted by
\begin{align}
\Y=  \mathcal{P}( \A\X+\blambda). \label{Eq:PoissonTransformationDefinition}
\end{align}

\section{Main Results} 
\label{sec:MainResults}
This section presents our main result pertaining to the linearity properties of the conditional expectation $ \E[ \X | \Y=\boldsymbol{y}]$. Specifically, our interest lies in answer various questions of optimality of linear estimators such as:
\begin{enumerate}
\item Under what prior distribution on $\X$ are linear estimators optimal for squared error loss and Bregman divergence\footnote{ Let $\phi: \Omega \to \mathbb{R}$ be a \emph{continuously-differentiable} and \emph{a strictly convex}  function defined on a \emph{closed convex} set $\Omega \subseteq \mathbb{R}^n$.   The Bregman divergence between $u$ and $v$, associated with the function $\phi$, is defined as
    $
        \ell_\phi(u,v) =\phi(u)-\phi(v)-\langle u-v, \nabla \phi(v) \rangle$. 
    } loss? Since the
conditional expectation is an optimal estimator for the aforementioned loss functions, this is equivalent to asking when 
the conditional expectation is a linear function of $\boldsymbol{y}$.  
\item  Which linear estimators are realizable from   $\E[\X| \Y] $? That is, given that  $\E[\X| \Y] =\HH \Y+\boldsymbol{c}$,  what values of the matrix $\HH$ and vector  $\boldsymbol{c}$ are permitted?
\item If the linear estimators are approximately optimal, can we say something about the prior distribution of $\X$? In other words, we
are looking for a quantitative refinement of 1).   
\end{enumerate}

Questions 1) and 2) are   answered in Theorem~\ref{thm:Main1} and Corollary~\ref{cor:InputXThm1}, and  question 3) is is addressed in Theorem~\ref{thm:QuantitativeRefinement}.

\subsection{Necessary and Sufficient Conditions for Linearity} 

Our first result is the following theorem, the proof of which can be found in Section~\ref{sec:thm:Main1}.

\begin{theorem}\label{thm:Main1}  Suppose that  $\Y=\mathcal{P}(\U)$ where $\U$ is a non-degenerate\footnote{A random vector is said to be degenerate of its covariance of matrix is not full rank. } random vector.   Then,  
\begin{align}
\E[\U| \Y=\boldsymbol{y}]=\HH \boldsymbol{y}+\boldsymbol{c},  \forall  \boldsymbol{y} \in   \mathbb{Z}^n_{+}   \label{eq:LinearityAssumption}
\end{align} 
if and only if
\begin{align}
P_{\U}=  \prod_{i=1}^n  \mathsf{Gam} \left(  \theta_i, \alpha_i \right). \label{eq:ProductGammaDistribuiton}
\end{align}
In this case
\begin{itemize}
    \item $\HH$ is diagonal with entries $\displaystyle h_{ii} = \frac{1}{1+\theta_i}$
    \item $\displaystyle c_i = \alpha_i h_{ii} = \frac{\alpha_i}{1+\theta_i}$
\end{itemize}
Note that $0 < h_{ii} < 1$ and $c_i > 0$ for all $i \in [1:n]$.
\end{theorem}

\subsection{Quantitative Refinement of Theorem~\ref{thm:Main1} } 

In this section, a quantitative refined of Theorem~\ref{thm:Main1} is shown. Namely, it is shown that if the conditional mean estimator is close to a linear function in a mean squared error sense, then the prior distribution must be close to a product gamma distribution in terms of their  characteristic functions. 

\begin{theorem}\label{thm:QuantitativeRefinement}   Let $\HH$ and $\boldsymbol{c}$ be as in Theorem~\ref{thm:Main1} and   let $\phi_{\mathsf{G}}$ denote the characteristic function of the product gamma distribution in \eqref{eq:ProductGammaDistribuiton}. 
Assume that  $\Y=\mathcal{P}(\U)$ for some $\U \in \mathbb{R}^n_{+}$   and that
\begin{align}
 \E \left[   \left \|\E[\U|\Y] -(\HH \Y+\boldsymbol{c}) \right \|^2 \right] \le \epsilon
\end{align}  
for some $\epsilon \ge 0$. 
Then,
\begin{align}
  \sup_{\boldsymbol{t}  \in \mathbb{R}^n}  \frac{|   \phi_{\U}(\boldsymbol{t} )   -  \phi_{\mathsf{G}}(\boldsymbol{t} )   |  }{ \|\boldsymbol{t} \| } \le \frac{\sqrt{ \epsilon}}{ 1- \max_{k} h_{kk}} , \label{eq:ControllOfCharPoisson}
\end{align} 
where $\phi_{\U}(\boldsymbol{t} )$ is the characteristic function of $\U$. 
\end{theorem} 
The proof of Theorem~\ref{thm:QuantitativeRefinement}  is presented in Section~\ref{proof:thm:QuantitativeRefinement}.

\section{Proof of Theorem~\ref{thm:Main1}}
\label{sec:thm:Main1}
We first establish conditions on $\boldsymbol{c}$ and $\HH$ under which the equality is possible. 

\subsection{Conditions $\boldsymbol{c}$} 
To establish such conditions we need the following representation of  the conditional expectation.  
\begin{lem} \label{lem:EmpericalBayes}  Let $P_\Y$ denote the probability mass function of $\Y$. Then, for $\boldsymbol{y} \in  \mathbb{Z}^n_{+}$
\begin{align}
\E[\U|\Y=\boldsymbol{y} ]  = \left(\diag(\boldsymbol{y})+\I  \right)  \frac{ \Delta P_\Y(\boldsymbol{y})}{ P_\Y(\boldsymbol{y})},  \label{eq:TRGformula}
\end{align} 
where
\begin{align}
[\Delta P_\Y(\boldsymbol{y})]_i= P_\Y(\boldsymbol{y}+\boldsymbol{1}_{i} ), \, i \in [1:n].
\end{align} 
\end{lem} 
The scalar version of Lemma~\ref{lem:EmpericalBayes} has been  shown in \cite{good1953population}  and in \cite{robbins1956empirical} and the vector version has been shown in \cite[Lemma~3]{palomar2007representation}  and  \cite[Lemma~3]{wang2014bregman}.

We proceed to show that every element of $\boldsymbol{c}$ must be strictly positive.   Choosing $\boldsymbol{y}=\boldsymbol{0}$ and combining \eqref{eq:LinearityAssumption} with  \eqref{eq:TRGformula}  implies that
\begin{align}
\boldsymbol{c} =   \frac{ \Delta P_\Y(\boldsymbol{0} )}{ P_\Y(\boldsymbol{0} )} ,
\end{align}
or equivalently for all $i$
\begin{align}
c_i =\frac{ P_\Y(\boldsymbol{0} + \boldsymbol{1}_{i} )}{ P_\Y(\boldsymbol{0} )}= \frac{   \E \left[  U_i \eu^{ -  \sum_{ i=1}^n U_i}   \right]    }{ \E \left[ \eu^{ -  \sum_{ i=1}^n U_i}   \right] }. 
\end{align} 
The above is zero if and only if $U_i=0$ and is positive otherwise.

\subsection{Conditions on $\HH$} 

We now proceed to study properties of $\HH$.  First, by combining \eqref{eq:LinearityAssumption} with  \eqref{eq:TRGformula}, we have
\begin{align}
 \frac{ \Delta P_\Y(\boldsymbol{y})}{ P_\Y(\boldsymbol{y})} &=   \left(\diag(\boldsymbol{y})+\I  \right)^{-1} (  \HH \boldsymbol{y}+\boldsymbol{c})  \\
 &=  \left(\diag(\boldsymbol{y})+\I  \right)^{-1}  \HH \boldsymbol{y}+   \left(\diag(\boldsymbol{y})+\I  \right)^{-1} \boldsymbol{c},
\end{align} 
which equivalently can be written as
\begin{align}
\frac{P_\Y(\boldsymbol{y} + \boldsymbol{1}_{i}  )}{P_\Y(\boldsymbol{y})}   = \frac{1}{y_i +1 } \sum_{j=1} h_{ij} y_j  +\frac{c_i}{y_i+1}, \forall i \in [1:n].  \label{eq:IdenityFOrEqch Term}
\end{align} 
Observe that every entry of  $ \frac{ \Delta P_\Y(\boldsymbol{y})}{ P_\Y(\boldsymbol{y})} $ is non-negative. Therefore, for every $i$ we have the following inequality:
\begin{align}
0 \le  \frac{1}{y_i +1 } \sum_{j=1} h_{ij} y_j  +\frac{c_i}{y_i+1}, \forall \boldsymbol{y}  \in   \mathbb{Z}^n_{+}, \label{eq:InequalitiesOnH}
\end{align} 
where $h_{ij}$ is the $(i,j)$  element of $\HH$.  Since $\boldsymbol{y}$ can be chosen arbitrary in \eqref{eq:InequalitiesOnH},  taking   limits along all possible paths as $y_i$'s go to infinity we arrive at
\begin{align}
0 \le h_{ii} +  \sum_{j \in S } h_{ij},   \forall i \,  \text{ and } \forall  S \subset[1:n]\setminus i .
\end{align} 
In particular, by selecting $S$ to be an empty set we arrive at the conclusion that $0 \le h_{ii}, \forall i$.   To see that $h_{ii} \neq 0$, consider 
\begin{align}
\E[U_i | \Y=\boldsymbol{0}+ y_i\boldsymbol{1}_{i}  ] =   h_{ii} y_i +c_i, \forall y \in \mathbb{Z}_{+}. 
\end{align} 
Therefore,   $ h_{ii}$ can only be zero if $U_i$ is a constant.  

Next, using  \eqref{eq:IdenityFOrEqch Term} and summing over $y_i$ we have that 
\begin{align}
 \sum_{y_i=0}^{k} (y_i+1)  P_{\Y}( \boldsymbol{y}+\boldsymbol{1}_{i}) 
 =   \sum_{y_i=0}^{k}    \left(  \sum_{j=1} h_{ij} y_j  +c_i  \right) P_{\Y}(\boldsymbol{y}),  \label{eq:TransitionEquation}
\end{align} 
or, equivalently,  by doing a change of variable on the left side of \eqref{eq:TransitionEquation},
\begin{align}
&\E[   Y_i 1_{ \{  Y_i \le k+1\}} | \Y_{-i}=\boldsymbol{y}_{-i}  ]  \notag\\
&= \E \left[ \left( \sum_{j=1} h_{ij} Y_j  +c_i  \right) 1_{ \{  Y_i \le k\}}   | \Y_{-i}=\boldsymbol{y}_{-i} \right] ,
\end{align} 
where $\Y_{-i}$ is $\Y$ with the $i$-th element removed. 
Now  by choosing $\boldsymbol{y}_{-i} =\boldsymbol{0} $ and re-arranging the terms we have that 
\begin{align}
h_{ii} &=  \frac{ \E[   Y_i 1_{ \{  Y_i \le k+1\}} | \Y_{-i}=\boldsymbol{0} ]  - c_i \E \left[ 1_{ \{  Y_i \le k\}}   | \Y_{-i}=\boldsymbol{0} \right]  }{\E[   Y_i 1_{ \{  Y_i \le k\}} | \Y_{-i}=\boldsymbol{0}   ]  } ,
\end{align} 
for all $k$. 
Now taking $k$ to infinity and using  the fact that $c_i>0$, it immediately follows that $h_{ii} <1$.      

The above discussion shows that $ 0< h_{ii} <1, \forall i$.    We now proceed to show that $\HH$ is invertible. To that end, we need the following lemma shown in Appendix~\ref{app:lem:ConditionalCovMatrix}. 

\begin{lem}\label{lem:ConditionalCovMatrix} For  $\boldsymbol{y} \in  \mathbb{Z}^n_{+}$ 
\begin{align}
&[ \mathsf{\boldsymbol{Var}}(\U|\Y=\boldsymbol{y}) ]_{ij}  \notag\\
&=   \E[ U_i | \Y= \boldsymbol{y} ]   \left(  \E[ U_j | \Y= \boldsymbol{y}+\boldsymbol{1}_{i} ]    -  \E[ U_j | \Y= \boldsymbol{y} ]   \right) .
\end{align}  
\end{lem} 

Now by using Lemma~\ref{lem:ConditionalCovMatrix} and  taking  $ \E[ \U| \Y= \boldsymbol{y} ] =\HH  \boldsymbol{y} +\boldsymbol{c}$ we have that 
\begin{align}
\mathsf{\boldsymbol{Var}}(\U|\Y=\boldsymbol{0})=  \boldsymbol{c}   \mathsf{1}^T  \odot \HH^T 
\end{align} 
where $\odot$ denotes the element-wise product (i.e., Hadamard product).  Now using an elementary rank  bound for the element-wise product,  and the fact that for non-degenerate random vectors $\mathsf{\boldsymbol{Var}}(\U|\Y=\boldsymbol{0})$ is a positive definite matrix, we have that 
\begin{align}
n& =\mathsf{Rank} \left( \mathsf{\boldsymbol{Var}}(\U|\Y=\boldsymbol{0})  \right) \\
&= \mathsf{Rank} \left(  \boldsymbol{c}   \mathsf{1}^T  \odot \HH^T  \right) \\
& \le  \mathsf{Rank} \left(  \boldsymbol{c}   \mathsf{1}^T  \right)  \mathsf{Rank} \left(    \HH^T  \right) \\
&=  \mathsf{Rank} \left(     \HH^T  \right) \\
& \le n.
\end{align}  
Therefore,  $\HH$ has  full rank and is invertible.

We now proceed to show that $\HH$ must be a diagonal matrix.  In order to that, we need the following definition.
\begin{definition}
The Laplace transform of the distribution of a random vector $\U  \in \mathbb{R}^n$  is denoted by
\begin{align}
\mathcal{L}_\U(\boldsymbol{t})= \E \left[ \eu^{-\boldsymbol{t}^T  \U } \right],  \boldsymbol{t}   \in \mathbb{R}^n_+. 
\end{align} 
\end{definition}

The following lemma is extensively used in this proof and the proof of Theorem~\ref{thm:QuantitativeRefinement}. 

\begin{lem} \label{lem:Gradient}  Let $\Y=\mathcal{P}(\U)$ and   suppose that \eqref{eq:LinearityAssumption} holds. Then,
\begin{align}
\E \left[  \left(\U -(\HH \Y+\boldsymbol{c}) \right) \eu^{-\boldsymbol{t} ^T\Y} \right] =\boldsymbol{0}  \label{eq:Consequence:Of:Orthogonality}
\end{align} 
for all $\boldsymbol{t}  \in \mathbb{R}^n_{+}$. 
Moreover,  for any $\boldsymbol{t}  \in \mathbb{R}^n_{+}$
\begin{align}
&\E \left[  \left(\U -(\HH \Y+\boldsymbol{c}) \right) \eu^{-\boldsymbol{t} ^T\Y} \right]   \notag\\
&  =  -  ( \HH (     \diag( {\bf s}  )- \I   )   + \I )  \nabla_{\bf s} \mathcal{L}_U({\bf s})-\boldsymbol{c}\mathcal{L}_U({\bf s}),
\label{eq:OrthgonalityIdenity}
\end{align}
where $ s_m=1-\eu^{-t_m}, m=1,\ldots,n$. 
\end{lem}
\begin{IEEEproof}
The proof of \eqref{eq:Consequence:Of:Orthogonality} follows from the orthogonality principle. 
To show \eqref{eq:OrthgonalityIdenity} we need to compute the following terms:
\begin{align}
\E \left[  \U \eu^{-\boldsymbol{t}^T\Y} \right],  \E \left[   \eu^{-\boldsymbol{t}  ^T\Y} \right] \text{ and }\E \left[ \Y  \eu^{-\boldsymbol{t} ^T\Y} \right]. 
\end{align} 
Also, recall that the Laplace transform of a distribution of a scalar Poisson random variable $W$ with the parameter $\lambda$ is given by
\begin{align}
\mathcal{L}_W(t)=\eu^{\lambda v(t)},
\end{align} 
where $v(t)=(\eu^{-t}-1)$. 

Now, first, 
\begin{align}
\E \left[  \U \eu^{\boldsymbol{t}^T\Y} \right]&=\E \left[  \U  \E\left[ \eu^{\boldsymbol{t}  ^T\Y}  \mid \U \right ] \right]\\
&=\E \left[  \U   \prod_{m=1}^n  \E\left[ \eu^{ t_m Y_m}  \mid U_m \right ] \right]\\
&=\E \left[  \U   \prod_{m=1}^n  \eu^{  v(t_m) U_m } \right] \label{eq:PoissonChar}\\
&=\E \left[  \U    \eu^{- {\bf s}^T \U  } \right]\\
&=  \nabla_{\bf s}  \mathcal{L}_U({\bf s}) , \label{eq:CharOfU}
\end{align} 
where \eqref{eq:PoissonChar} follows by using the Laplace transform of a scalar Poisson distribution.  Second,  using similar steps, we have 
\begin{align}
\E \left[   \eu^{-\boldsymbol{t}  ^T\Y} \right]&= \E \left[  \E\left[ \eu^{-\boldsymbol{t}  ^T\Y}  \mid \U \right ] \right]\\
&=\E \left[   \prod_{i=m}^n  \eu^{  v(t_m) U_i } \right]\\
&=\E \left[   \eu^{ -{\bf s}^T \U  } \right]\\
&= \mathcal{L}_U({\bf s}). \label{eq:CharY}
\end{align} 
Third,  
\begin{align}
\E \left[ \Y  \eu^{-\boldsymbol{t} ^T\Y} \right]&=  -\nabla_{\boldsymbol{t} }  \E \left[  \eu^{-\boldsymbol{t} ^T\Y} \right]\\
&= -\nabla_{\boldsymbol{t} }  \E \left[   \prod_{m=1}^n  \eu^{  v(t_m) U_m } \right]\\
&= -\nabla_{\boldsymbol{t} }  \E \left[   \eu^{ - {\bf s}^T \U  } \right]\\
&=   \E \left[ \nabla_{\boldsymbol{t} } {\bf s}^T \U   \eu^{ -{\bf s}^T \U  } \right]\\
&=   \E \left[  (  \I-   \diag( {\bf s}  ) ) \U    \eu^{ - {\bf s}^T \U  }  \right]\\
&=    (     \diag( {\bf s}  )- \I  )   \nabla_{\bf s} \mathcal{L}_U({\bf s}),  \label{eq:GradOfYchar}
\end{align}
where we have used that 
\begin{align}
\frac{{\rm d}}{ {\rm d} t_m } s_m U_m&=\frac{{\rm d}}{ {\rm d} t_m }  (1- \eu^{-t_m}) U_m\\
&=   \eu^{-t_m} U_m\\
&=   (1-s) U_m.
\end{align} 

Combining \eqref{eq:CharOfU}, \eqref{eq:CharY} and \eqref{eq:GradOfYchar}  we arrive at
\begin{align}
&\E \left[  \left(\U -(\HH \Y+\boldsymbol{c}) \right) \eu^{-\boldsymbol{t}  ^T\Y} \right]  \\
&=  \nabla_{\bf s}  \mathcal{L}_U({\bf s})- \HH (     \diag( {\bf s}  )- \I  )   \nabla_{\bf s} \mathcal{L}_U({\bf s})-\boldsymbol{c}\mathcal{L}_U({\bf s})\\
&=  -  ( \HH      \diag( {\bf s}  )   + (\I-\HH) )  \nabla_{\bf s} \mathcal{L}_U({\bf s})-\boldsymbol{c}\mathcal{L}_U({\bf s}). 
\end{align} 
This concludes the proof. 
\end{IEEEproof}

To present  the solution to the differential equation in \eqref{eq:OrthgonalityIdenity} we need the following lemma. 

First using  that $\HH$ is invertible it follows that 
\begin{align} 
\frac{ \nabla_{\bf s} \mathcal{L}_{\bf U}({\bf s}) }{ \mathcal{L}_{\bf U}({\bf s})} = -  \left(    \HH^{-1} (\I- \HH) +   \diag ( {\bf s})     \right)^{-1}   \HH^{-1}  \boldsymbol{c}  ,
\end{align}  
which can further be simplified to 
\begin{align}
\nabla g( {\bf s})= \left(   \HH^{-1} (\I- \HH) +   \diag ( {\bf s})     \right)^{-1}   \HH^{-1}  \boldsymbol{c} , \label{eq:SimplificationOfmatrixPDF}
\end{align} 
where  $g( {\bf s})=  \log (\mathcal{L}_{\bf U}({\bf s}) )$. 

Next it is shown that \eqref{eq:SimplificationOfmatrixPDF} has a solution only if $\HH$ is a diagonal matrix and the solution is characterized.  

\begin{lem}\label{lem:SolutionToDifferentialEquation} For $ \boldsymbol{0}  \prec \A \in \mathbb{R}^{n \times n}$ and ${\bf b} \in \mathbb{R}^n$ where ${\bf b}$ is assumed to have all positive entries.   The system 
\begin{align}
\nabla g( {\bf s})=- ( \A + \diag ( {\bf s}) )^{-1} {\bf b}, \,  g( \boldsymbol{0}  )=\boldsymbol{0} , \label{eq:SystemWeHaveTOSolve}
\end{align} 
has a solution only if   $\A $ is a diagonal matrix with a solution given by 
\begin{align}
g( {\bf s})& = \sum_{i=1}^n b_i \log \left( 1+ \frac{s_i}{A_{ii}} \right). 
\end{align}  
\end{lem} 
\begin{IEEEproof} 
We first find the Hessian matrix  of $f({\bf s})=\nabla g( {\bf s})$. Let
\begin{align}
\C&= \A + \diag ( {\bf s}),\\
 \boldsymbol{\mathsf{S}}&= \diag ( {\bf s}), \\
 \boldsymbol{\mathsf{S}} f&=   \diag (f)  {\bf s}.
\end{align} 
Then, the differential is given by 
\begin{align}
\partial  f  &=\partial  \C^{-1} {\bf b}\\
&=- \C^{-1}  (\partial  \C )   \C^{-1} {\bf b}\\
&=- \C^{-1}  (\partial  \C )   f\\
&=- \C^{-1}  (\partial  \boldsymbol{\mathsf{S}} )   f\\
&=- \C^{-1}  \diag (f)  \partial  {\bf s}. 
\end{align}
Hence,
\begin{align}
\frac{\partial f}{ \partial  {\bf s}}=  - \C^{-1}  \diag (f)= - \C^{-1}  \diag (  ( \A + \diag ( {\bf s}) )^{-1} {\bf b} ). 
\end{align} 
Therefore, the Hessian matrix of  $g$ is given by 
\begin{align}
\nabla^2 g( {\bf s}) &= -  ( \A + \diag ( {\bf s}) )^{-1}   \diag \left( ( \A + \diag ( {\bf s}) )^{-1}   {\bf b}   \right).  \label{eq:HessianMatrix}
\end{align}
Note that the Hessian matrix must be  symmetric.  Next, it is shown that in order for the Hessian to be symmetric $\A$ must be a diagonal matrix.  

Let $\tilde{\A} =  ( \A + \diag ( {\bf s}) )^{-1}  $ and choose ${\bf s}$ such that  
\begin{align}
\tilde{{\bf b}}= \tilde{\A} {\bf b}=  ( \A + \diag ( {\bf s}) )^{-1} {\bf b}
\end{align}  has distinct elements all of which are non-zero.  Note that this is possible in view of the assumption that   ${\bf b}$ has non-zero entries. 

Next, observe that if  $\tilde{\A} \diag(\tilde{{\bf b}})$ is symmetric, then  $\tilde{\A} $ must be symmetric. This follows by letting  $  \tilde{\C}= \tilde{\A} \diag(\tilde{{\bf b}})$   and observing that $\tilde{\A}= \tilde{\C}\diag(\tilde{{\bf b}})^{-1}$ is symmetric.  The symmetry of  $\tilde{\A}$ implies that 
\begin{align}
\tilde{\A} \diag(\tilde{{\bf b}})=  \diag(\tilde{{\bf b}}) \tilde{\A}^{T}=  \diag(\tilde{{\bf b}}) \tilde{\A}.
\end{align}
In other words,  $\tilde{\A}$ and $ \diag(\tilde{{\bf b}})$ commute. However, if all elements of a diagonal matrix are distinct, then it commutes only with a diagonal matrix.  Therefore,    $\tilde{\A}$ is a diagonal matrix.    This implies that for the Hessian to be symmetric $\A $ must be a diagonal matrix. 

Since $\A$ is diagonal, the solution is obtained by an application of the   fundamental theorem of calculus for line integrals: for a function $f$ and a smooth curve $\boldsymbol{r}(t)$ we have 
\begin{align}
\int_{a}^b \nabla f(\boldsymbol{r}(t)) \boldsymbol{\cdot} \dot{\boldsymbol{r}}(t) {\rm d} t= f (\boldsymbol{r}(b))-  f (\boldsymbol{r}(a)).  \label{eq:FTCm}
\end{align} 
Applying \eqref{eq:FTCm} to \eqref{eq:SystemWeHaveTOSolve} with a choice of $\boldsymbol{r}(t)= (1-t) \boldsymbol{0}  + t {\bf s}, \, t\in (0,1)$, we have that 
\begin{align}
g({\bf s}) &=-   \int_0^1    \left(          \A +\diag( {\bf s})t   \right)^{-1} {\bf b} \cdot  {\bf s}   {\rm d} t\\
&=-   {\bf s}^T \int_0^1      \left(      \A +\diag( {\bf s})t    \right)^{-1}     {\rm d} t {\bf b}\\
&=-   {\bf s}^T  \diag \left( \left[  \frac{\log( 1+ \frac{s_k}{A_{kk}} )}{s_{k}    }   \right]_k \right)   {\bf b}\\
&= -  \sum_{k=1}^n b_k \log \left( 1+ \frac{s_k}{A_{kk}} \right). 
\end{align} 
\end{IEEEproof} 
Setting $\A= \HH^{-1} (\I-\HH)$ and $ {\bf b} =\HH^{-1} \boldsymbol{c}$ in Lemma~\ref{lem:SolutionToDifferentialEquation}  and using that  $g( {\bf s})=  \log (\mathcal{L}_{\bf U}({\bf s}) )$ we arrive at the following form for the Laplace transform of  the distribution of $\U$: 
\begin{align}
\mathcal{L}_{\bf U}({\bf s}) =  \prod_{k=1}^n  \frac{1}{\left( 1+ \frac{ h_{kk}s_k}{1- h_{kk}} \right)^{  \frac{h_{kk}}{c_k} }},
\end{align} 
which is the Laplace transform of a product of   Gamma distributions.

\section{Proof Theorem~\ref{thm:QuantitativeRefinement}} 
\label{proof:thm:QuantitativeRefinement}

Let the characteristic function of the product gamma distribution be denoted by
\begin{align}
\phi_{\mathsf{G}}(\boldsymbol{t} )= \prod_{k=1}^n  \left(1 -\frac{i  t_k}{\alpha_k}  \right)^{-\theta_k}. 
\end{align} 
The following result,  which is a generalization of the scalar result in \cite{dytso2019estimation}, will be useful. 

\begin{lem}\label{lem:BoundOnCharFunc} Let $ \phi_{\U}(\boldsymbol{t} ) $ be a characteristic function of  a distribution of a non-negative random vector $\U$ and let
\begin{align}
\A&= \diag^{-1} \left ( [\alpha_1, \ldots, \alpha_n]^T \right),  \\
\tilde{\boldsymbol{c}}&=   \left  [ \frac{\theta_1}{\alpha_1}, \ldots,  \frac{\theta_k}{\alpha_k} \right]^T    .
\end{align} 
Then,  for every $ \boldsymbol{t} \in \mathbb{R}^n$
\begin{align}
& |   \phi_{\U}(\boldsymbol{t} )   -  \phi_{\mathsf{G}}(\boldsymbol{t} )   |  \notag\\
 & \le   \|\boldsymbol{t} \|  \sup_{ \boldsymbol{t} \in\mathbb{R}^n}  \left \|    \left(i\I + \A \diag( \boldsymbol{t})  \right)  \nabla \phi_{\U}( \boldsymbol{t})   + \tilde{\boldsymbol{c}}  \phi_{\U}( \boldsymbol{t})    \right\|  . 
\end{align} 
\end{lem} 
\begin{IEEEproof}
First, note that
\begin{align}
\frac{\partial}{\partial t_k}  \frac{1}{\phi_{\mathsf{G}}(\boldsymbol{t} ) }= - \frac{i \theta_k}{\alpha_k}   \frac{1 }{ \left(1 -\frac{i t_k}{\alpha_k} \right)  \phi_{\mathsf{G}}(\boldsymbol{t} )} , 
\end{align} 
and hence
\begin{align}
&\frac{\partial}{\partial t_k}  \phi_{\U}(\boldsymbol{t} ) \frac{1}{\phi_{\mathsf{G}}(\boldsymbol{t})} \notag\\
&=  \frac{\partial}{\partial t_k}  \phi_{\U}(\boldsymbol{t} ) \frac{1}{\phi_{\mathsf{G}}(\boldsymbol{t})} +   \phi_{\U}(\boldsymbol{t} )  \frac{\partial}{\partial t_k} \frac{1}{\phi_{\mathsf{G}}(\boldsymbol{t})}\\
&=   \frac{1 }{ \left(1 -\frac{i t_k}{\alpha_k} \right)  \phi_{\mathsf{G}}(\boldsymbol{t} )} \left(   \left(1 -\frac{i t_k}{\alpha_k} \right)  \frac{\partial}{\partial t_k}  \phi_{\U}(\boldsymbol{t} )   - \frac{i \theta_k}{\alpha_k}  \phi_{\U}(\boldsymbol{t} )  \right) \\
&=   \frac{-i }{ \left(1 -\frac{i t_k}{\alpha_k} \right)  \phi_{\mathsf{G}}(\boldsymbol{t} )} \left(   \left(i +\frac{ t_k}{\alpha_k} \right)  \frac{\partial}{\partial t_k}  \phi_{\U}(\boldsymbol{t} )   + \frac{ \theta_k}{\alpha_k}  \phi_{\U}(\boldsymbol{t} )  \right) . 
\end{align} 

Therefore, the gradient can be upper bounded as 
\begin{align}
& \left \| \nabla\left(  \phi_{\U}(\boldsymbol{t} ) \frac{1}{\phi_{\mathsf{G}}(\boldsymbol{t})} \right) \right\| \notag\\
 & = \frac{  \left \|  \left( \I- i  \A \diag(\boldsymbol{t} ) \right)^{-1} \left(    \left(i\I + \A \diag(\boldsymbol{t} )  \right)  \nabla \phi_{\U}(\boldsymbol{t} )   + \tilde{\boldsymbol{c}}  \phi_{\U}(\boldsymbol{t} )  \right)  \right\|}{ |\phi_{\mathsf{G}}(\boldsymbol{t} )   |  } \\
 & \le  \hspace{-0.05cm} \frac{  \left \|  \left( \I- i  \A \diag(\boldsymbol{t} ) \right)^{-1} \right \|_{*}  \hspace{-0.05cm} \left \|   \hspace{-0.03cm}  \left(i\I + \A \diag(\boldsymbol{t} )  \right)  \hspace{-0.05cm} \nabla \phi_{\U}(\boldsymbol{t} ) \hspace{-0.05cm}   + \hspace{-0.05cm} \tilde{\boldsymbol{c}}  \phi_{\U}(\boldsymbol{t} )   \hspace{-0.03cm}  \right\|}{ |\phi_{\mathsf{G}}(\boldsymbol{t} )   |  }  \label{eq:BoundOnGradient}
\end{align} 
where  $\| \cdot \|_{\star}$ denotes the operator norm.

 Next, recall that the operator norm of a  diagonal matrix is given by the maximal element  and 
\begin{align}
\left \|  \left( \I- i  \A \diag(\boldsymbol{t} ) \right)^{-1} \right \|_{*}  &= \max_{k \in [1:n] }  \left| 1 - i \frac{t_k}{\alpha_k}  \right|^{-1}\\
&=  \frac{1}{\sqrt{ 1 +    \min_{k \in [1:n] }   \frac{t_k^2}{\alpha_k^2} }}. 
\end{align} 

Moreover, note that
\begin{align}
 |\phi_{\mathsf{G}}(\boldsymbol{t} )   | = \left|  \prod_{i=1}^n  \left(1 -\frac{i  t_i}{\alpha_i}  \right)^{-\theta_i}  \right| =   \prod_{i=1}^n  \left ( 1 +\frac{  t_i^2}{\alpha_i^2}   \right) ^{- \frac{\theta_i}{2}} . 
\end{align}

Now let $r(\tau)= \tau \boldsymbol{t}$ and observe the following sequence of steps: 
\begin{align}
& |   \phi_{\U}(\boldsymbol{t} )   -  \phi_{\mathsf{G}}(\boldsymbol{t} )   | \notag\\
 &=     | \phi_{\mathsf{G}}(\boldsymbol{t} ) | \left|   \frac{ \phi_{\U}(\boldsymbol{t} ) }{ \phi_{\mathsf{G}}(\boldsymbol{t} ) }  -  1 \right |   \\
 &= | \phi_{\mathsf{G}}(\boldsymbol{t} ) |  \left |  \int_0^1 \nabla  \frac{ \phi_{\U}(  \boldsymbol{r}(\tau) ) }{ \phi_{\mathsf{G}}( \boldsymbol{r}(\tau)) }  \boldsymbol{\cdot} \dot{\boldsymbol{r}}(\tau)  {\rm d} \tau    \right |  \label{eq:FTCapp:secThm} \\
 & \le    | \phi_{\mathsf{G}}(\boldsymbol{t} ) |   \int_0^1  \left\| \nabla  \frac{ \phi_{\U}(  \boldsymbol{r}(\tau) ) }{ \phi_{\mathsf{G}}( \boldsymbol{r}(\tau)) } \right\|  \left \|\dot{\boldsymbol{r}}(\tau) \right\|  {\rm d} \tau   \label{eq:CSapplic:secThm}   \\
  &  \le  \|\boldsymbol{t} \|  \sup_{ \boldsymbol{t} \in\mathbb{R}^n}  \left \|    \left(i\I + \A \diag( \boldsymbol{t})  \right)  \nabla \phi_{\U}( \boldsymbol{t})   + \tilde{\boldsymbol{c}}  \phi_{\U}( \boldsymbol{t})    \right\|,    \label{eq:MaximumElementBound}
\end{align} 
where \eqref{eq:FTCapp:secThm}  follows from the fundamental theorem of calculus for line integrals;   \eqref{eq:CSapplic:secThm} follows by using the Cauchy-Schwarz inequality;  and  \eqref{eq:MaximumElementBound} follows by  using the bound in \eqref{eq:BoundOnGradient}, the fact that $ | \phi_{\mathsf{G}}(\tau\boldsymbol{t} ) | $ is an increasing function of $\tau$, and 
\begin{align}
& \int_0^1    \frac{  \left \|    \left(i\I + \A \diag( \tau\boldsymbol{t})  \right)  \nabla \phi_{\U}( \tau\boldsymbol{t})   + \tilde{\boldsymbol{c}}  \phi_{\U}( \tau\boldsymbol{t})    \right\|  }{ |\phi_{\mathsf{G}}(\tau \boldsymbol{t} )   |  \sqrt{ 1 +    \min_{k \in [1:n] }   \frac{ \tau^2 t_k^2}{\alpha_k^2} }} {\rm d} \tau \notag\\
& \le  \frac{1}{ |\phi_{\mathsf{G}}( \boldsymbol{t} )   |  } \int_0^1    \frac{  \left \|    \left(i\I + \A \diag( \tau\boldsymbol{t})  \right)  \nabla \phi_{\U}( \tau\boldsymbol{t})   + \tilde{\boldsymbol{c}}  \phi_{\U}( \tau\boldsymbol{t})    \right\|  }{ \sqrt{ 1 +    \min_{k \in [1:n] }   \frac{ \tau^2 t_k^2}{\alpha_k^2} }} {\rm d} \tau \notag\\
 & \le  \sup_{ \boldsymbol{t} \in\mathbb{R}^n}  \left \|    \left(i\I + \A \diag( \boldsymbol{t})  \right)  \nabla \phi_{\U}( \boldsymbol{t})   + \tilde{\boldsymbol{c}}  \phi_{\U}( \boldsymbol{t})    \right\|  .  
\end{align}  
This concludes the proof. 
\end{IEEEproof}

With Lemma~\ref{lem:BoundOnCharFunc} at our disposal we are now ready to proof the main result.  First, note that by using  a simple transformation from the Laplace transform to the characteristic function, the result in Lemma~\ref{lem:Gradient} can be re-written as
 \begin{align}
&\E \left[  \left(\U -(\HH \Y+\boldsymbol{c}) \right) \eu^{ i\boldsymbol{t} ^T\Y} \right] \notag\\
&=-  (   i(\I -\HH)+\HH     \diag( {\bf s}  )  )  \nabla_{\bf s} \phi_\U({\bf s})-\boldsymbol{c}\phi_\U({\bf s}). 
\end{align} 
Moreover,  
\begin{align}
& \left \|   (  i(\I -\HH)+\HH     \diag( {\bf s}  ) )  \nabla_{\bf s} \phi_\U({\bf s})+\boldsymbol{c}\phi_\U({\bf s}) \right\| \notag\\
 &= \left \| \E \left[  \left(\U -(\HH \Y+\boldsymbol{c}) \right) \eu^{ i\boldsymbol{t} ^T\Y} \right] \right \|\\
 &=   \left \| \E \left[  \left(\E[\U|\Y] -(\HH \Y+\boldsymbol{c}) \right) \eu^{ i\boldsymbol{t} ^T\Y} \right]  \right.  \notag\\
 & \left. \quad  + \E \left[  \left(\U -(\E[\U|\Y]  \right) \eu^{ i\boldsymbol{t} ^T\Y} \right] \right \| \label{eq:ApplicationOfOrthgonalityPrinciple}\\
  &=   \left \| \E \left[  \left(\E[\U|\Y] -(\HH \Y+\boldsymbol{c}) \right) \eu^{ i\boldsymbol{t} ^T\Y} \right] \right \|\\
  &\le   \E \left[   \left \|\E[\U|\Y] -(\HH \Y+\boldsymbol{c}) \right \| \right] \label{eq:ModulusInequalityAppliation} \\
   &\le  \sqrt{ \E \left[   \left \|\E[\U|\Y] -(\HH \Y+\boldsymbol{c}) \right \|^2 \right]},  \label{eq:J-app}
  \end{align}
  where \eqref{eq:ApplicationOfOrthgonalityPrinciple} follows by the orthogonality principle; \eqref{eq:ModulusInequalityAppliation} follows by using the modulus inequality; and \eqref{eq:J-app} follows by using Jensen's inequality.

Now by setting   $\tilde{\boldsymbol{c}}  = (\I - \HH)^{-1}  \boldsymbol{c}$ and  $\A=  (\I - \HH)^{-1} \HH$ in Lemma~\ref{lem:BoundOnCharFunc} we have that
\begin{align}
& |   \phi_{\U}(\boldsymbol{t} )   -  \phi_{\mathsf{G}}(\boldsymbol{t} )   |  \notag\\
 & \le   \|\boldsymbol{t} \|  \sup_{ \boldsymbol{t} \in\mathbb{R}^n}  \left \|    \left(i\I + \A \diag( \boldsymbol{t})  \right)  \nabla \phi_{\U}( \boldsymbol{t})   + \tilde{\boldsymbol{c}}  \phi_{\U}( \boldsymbol{t})    \right\|  \\
  & \le   \|\boldsymbol{t} \|   \| (\I -\HH)^{-1}  \|_{\star}    \notag\\
  &\quad \cdot \sup_{ \boldsymbol{t} \in\mathbb{R}^n}   \left \|   (  i(\I -\HH)+\HH     \diag( {\bf s}  ) )  \nabla_{\bf s} \phi_\U({\bf s})+\boldsymbol{c}\phi_\U({\bf s}) \right\| \\
  & \le   \|\boldsymbol{t} \|   \| (\I -\HH)^{-1}  \|_{\star}  \sqrt{ \E \left[   \left \|\E[\U|\Y] -(\HH \Y+\boldsymbol{c}) \right \|^2 \right]} \\
    & =   \|\boldsymbol{t} \|  \frac{\sqrt{ \epsilon}}{ 1- \max_{k} h_{kk}} .
\end{align} 
This concludes the proof.

\section{Concluding Remarks} 
\label{sec:Applications}

This section discusses implications of our results for the practically relevant model  $ \Y=  \mathcal{P}( \A\X+\blambda)$, which explicitly takes into account the intensity matrix $\A$ and the dark current parameter $\blambda$.     In addition, we also compare the Poisson result obtained in this work to their Gaussian counterparts. 

We begin by adopting  Theorem~\ref{thm:Main1} to the parametrization $ \Y=  \mathcal{P}( \A\X+\blambda)$.  This is done by setting $\U=\A\X+\blambda$ in Theorem~\ref{thm:Main1}.

\begin{corollary}\label{cor:InputXThm1} Suppose that $\Y=  \mathcal{P}( \A\X+\blambda)$. Then,    
\begin{align}
\E[\X| \Y=\boldsymbol{y}]=\C \boldsymbol{y}+  {\bf b},  \forall  \boldsymbol{y} \in   \mathbb{Z}^n_{+}  \label{eq:cor:linearAssumption}
\end{align} 
 if and only if    all of the following conditions hold:
\begin{itemize}
\item $\blambda=\boldsymbol{0} $;  
\item  $\A \C$ is a diagonal matrix  with  $ 0< [\A \C]_{ii}<1, \forall i \in [1:n]$;
\item $\A  {\bf b} $ is a vector of positive elements; and
\item $P_{\A \X}=  \prod_{i=1}^n  \mathsf{Gam} \left(  \frac{1-[\A \C]_{ii}}{[\A \C]_{ii}}, \frac{[ \A  {\bf b}  ]_{i}}{ [\A \C]_{ii} }\right)$
\end{itemize} 
\end{corollary}

\begin{IEEEproof}
Let $\U=\A\X+\blambda$. By multiplying \eqref{eq:cor:linearAssumption} by $\A$ and adding $\blambda$ we have that 
\begin{align}
\E[\U| \Y=\boldsymbol{y}]=  \A \E[ \X | \Y=\boldsymbol{y}] +\blambda=   \A\C \boldsymbol{y}+  \A{\bf b} +\blambda. 
\end{align} 
Next, note that the linearity of the conditional expectation implies that $\U=\A\X+\blambda$ is according with a product gamma distribution which has  non-negative support.  However,  if $\blambda $ has  positive components, this would imply that $\A \X=\U-\blambda $ has negative components, which is not allowed under the Poisson model.  Therefore, $\blambda $ must be zero.  

The rest of the argument follows  from Theorem~\ref{thm:Main1} by mapping  $\A\C$ to $\HH$ and  $ \A{\bf b}  $ to $\boldsymbol{c}$. 
\end{IEEEproof} 

A few comments are now in order. 

\subsection{The case of a non-zero dark current} 

Somewhat regrettably Corollary~\ref{cor:InputXThm1} shows that the conditional expectation can only be linear if the dark current parameter is zero.   To demonstrate the effect of the dark current we  investigate a scalar case with an  exponential distribution as a prior (i.e., a gamma distribution with $\theta=1$). 

\begin{lem}\label{lem:OutputExponential}  Let $Y=\mathcal{P}(aX+\lambda)$ and take  $X$ to be an exponential random variable of rate $\alpha$. Then, for every $a>0$ and $\lambda \ge 0$ 
\begin{align}
\E[X|Y=k]=  \frac{1}{a}  \frac{ (k+1) P_Y(k+1)}{P_Y(k)}-\frac{\lambda}{a},  \label{eq:expressionForConditional}
\end{align}
where
\begin{align}
&P_Y(0)=  \frac{\alpha \eu^{-\lambda}}{ \alpha+a},   \label{eq:OutputPMF0}  \\
&P_Y(k)=\frac{\Gamma(k+1,\lambda)}{\Gamma(k+1)}-\frac{\Gamma(k,\lambda)}{\Gamma(k)} \notag\\
&+\frac{\eu^{\frac{\alpha}{a} \lambda}}{ \left( 1+\frac{\alpha}{a}  \right)^{k}} \left(   \frac{\Gamma \left( k, \lambda \left(\frac{\alpha}{a} +1 \right) \right)  }{\Gamma \left( k \right)   }  -     \frac{\Gamma \left( k+1, \lambda \left(\frac{\alpha}{a} +1 \right) \right)  }{\Gamma \left( k+1 \right)   \left( 1+\frac{\alpha}{a}  \right) }   \right),  \label{eq:OutputPMF1} 
\end{align}
\end{lem}
where  $\Gamma(\cdot, \cdot)$ is the upper incomplete gamma function. 
\begin{IEEEproof}
\eqref{eq:expressionForConditional} is a scalar version of Lemma~\ref{lem:EmpericalBayes}. The proof of \eqref{eq:OutputPMF0} and \eqref{eq:OutputPMF1} follows by invoking standard integration techniques for exponential functions. 
\end{IEEEproof}

 The effect of the dark current parameter on the conditional expectation in the scalar case   for an exponential random variable is shown in Fig.~\ref{fig:ConditionalExpectationExponential}.  Observe that the larger the dark current, the smaller the conditional expectation is.  The interpretation here is that large values of dark current inflate the observed count at $Y$, and the estimator compensates by producing smaller estimates of $X$. 

It is also interesting to compare the optimal linear estimator under the squared error loss to the conditional expectation. The former is given by 
\begin{align}
\widehat{X}(y)&= c y +b,\\
c&=   \frac{a \mathbb{V}(X)}{a^2  \mathbb{V}(X)  + a\E[X] +\lambda },\\
b&= \E[X] -c (a \E[X]+\lambda).
\end{align} 

\begin{figure}[h!]  
	\centering   
%
%
\begin{tikzpicture}

\begin{axis}[%
width=7.4cm,
height=5.7cm,
at={(1.159in,0.77in)},
scale only axis,
xmin=0.656682027649771,
xmax=35,
xlabel style={font=\color{white!15!black}},
xlabel={$k$},
ymin=0,
ymax=11.1,
ylabel style={font=\color{white!15!black}},
ylabel={$\E[X|Y=k]$},
axis background/.style={fill=white},
xmajorgrids,
ymajorgrids,
legend pos=north west,
legend style={legend cell align=left, align=left, draw=white!15!black, font=\footnotesize}
]

\addplot [color=black,  dashed,  line width=0.3mm]
  table[row sep=crcr]{%
0	0.25\\
1	0.277777777777777\\
2	0.310975609756097\\
3	0.350923482849605\\
4	0.399270482603816\\
5	0.458016606244889\\
6	0.529503770487892\\
7	0.616329399926177\\
8	0.721140522247445\\
9	0.846281655353462\\
10	0.993322128504657\\
11	1.16257664098021\\
12	1.35281277539743\\
13	1.56132926734424\\
14	1.78444178471602\\
15	2.01820177922015\\
16	2.25905964659576\\
17	2.50425438672973\\
18	2.75188902355175\\
19	3.00079420035606\\
20	3.25031847529539\\
21	3.5001374695329\\
22	3.74996857025491\\
23	4.00005617290663\\
24	4.25001957522359\\
25	4.50000653640201\\
26	4.75000209499854\\
27	5.00000064556972\\
28	5.25000019154265\\
29	5.50000005479636\\
30	5.75000001513423\\
31	6.00000000404028\\
32	6.25000000104374\\
33	6.50000000026119\\
34	6.75000000006338\\
35	7.00000000001492\\
36	7.25000000000342\\
37	7.50000000000076\\
38	7.75000000000016\\
39	8.00000000000004\\
40	8.25000000000001\\
41	8.5\\
42	8.75\\
43	9\\
44	9.25\\
45	9.5\\
46	9.75\\
47	10\\
48	10.25\\
49	10.5\\
50	10.75\\
51	11\\
52	11.25\\
53	11.5\\
54	11.75\\
55	12\\
56	12.25\\
57	12.5\\
58	12.75\\
59	13\\
};
\addlegendentry{Conditional Expectation for $\lambda=2$}

\addplot [color=red,  line width=0.3mm]
  table[row sep=crcr]{%
0	0.227272727272727\\
1	0.272727272727273\\
2	0.318181818181818\\
3	0.363636363636364\\
4	0.409090909090909\\
5	0.454545454545454\\
6	0.5\\
7	0.545454545454545\\
8	0.590909090909091\\
9	0.636363636363636\\
10	0.681818181818182\\
11	0.727272727272727\\
12	0.772727272727273\\
13	0.818181818181818\\
14	0.863636363636364\\
15	0.909090909090909\\
16	0.954545454545454\\
17	1\\
18	1.04545454545455\\
19	1.09090909090909\\
20	1.13636363636364\\
21	1.18181818181818\\
22	1.22727272727273\\
23	1.27272727272727\\
24	1.31818181818182\\
25	1.36363636363636\\
26	1.40909090909091\\
27	1.45454545454545\\
28	1.5\\
29	1.54545454545455\\
30	1.59090909090909\\
31	1.63636363636364\\
32	1.68181818181818\\
33	1.72727272727273\\
34	1.77272727272727\\
35	1.81818181818182\\
36	1.86363636363636\\
37	1.90909090909091\\
38	1.95454545454545\\
39	2\\
40	2.04545454545454\\
41	2.09090909090909\\
42	2.13636363636364\\
43	2.18181818181818\\
44	2.22727272727273\\
45	2.27272727272727\\
46	2.31818181818182\\
47	2.36363636363636\\
48	2.40909090909091\\
49	2.45454545454545\\
50	2.5\\
51	2.54545454545454\\
52	2.59090909090909\\
53	2.63636363636364\\
54	2.68181818181818\\
55	2.72727272727273\\
56	2.77272727272727\\
57	2.81818181818182\\
58	2.86363636363636\\
59	2.90909090909091\\
60	2.95454545454545\\
61	3\\
62	3.04545454545454\\
63	3.09090909090909\\
64	3.13636363636364\\
65	3.18181818181818\\
66	3.22727272727273\\
67	3.27272727272727\\
68	3.31818181818182\\
69	3.36363636363636\\
70	3.40909090909091\\
71	3.45454545454545\\
72	3.5\\
73	3.54545454545454\\
74	3.59090909090909\\
75	3.63636363636364\\
76	3.68181818181818\\
77	3.72727272727273\\
78	3.77272727272727\\
79	3.81818181818182\\
80	3.86363636363636\\
81	3.90909090909091\\
82	3.95454545454545\\
83	4\\
84	4.04545454545454\\
85	4.09090909090909\\
86	4.13636363636364\\
87	4.18181818181818\\
88	4.22727272727273\\
89	4.27272727272727\\
90	4.31818181818182\\
91	4.36363636363636\\
92	4.40909090909091\\
93	4.45454545454545\\
94	4.5\\
95	4.54545454545454\\
96	4.59090909090909\\
97	4.63636363636364\\
98	4.68181818181818\\
99	4.72727272727273\\
100	4.77272727272727\\
};
\addlegendentry{Optimal Linear Estimator for $\lambda=2$}

\addplot [color=blue,  line width=0.3mm]
  table[row sep=crcr]{%
0	-1.75\\
1	-1.5\\
2	-1.25\\
3	-1\\
4	-0.75\\
5	-0.5\\
6	-0.25\\
7	0\\
8	0.25\\
9	0.5\\
10	0.75\\
11	1\\
12	1.25\\
13	1.5\\
14	1.75\\
15	2\\
16	2.25\\
17	2.5\\
18	2.75\\
19	3\\
20	3.25\\
21	3.5\\
22	3.75\\
23	4\\
24	4.25\\
25	4.5\\
26	4.75\\
27	5\\
28	5.25\\
29	5.5\\
30	5.75\\
31	6\\
32	6.25\\
33	6.5\\
34	6.75\\
35	7\\
36	7.25\\
37	7.5\\
38	7.75\\
39	8\\
40	8.25\\
41	8.5\\
42	8.75\\
43	9\\
44	9.25\\
45	9.5\\
46	9.75\\
47	10\\
48	10.25\\
49	10.5\\
50	10.75\\
51	11\\
52	11.25\\
53	11.5\\
54	11.75\\
55	12\\
56	12.25\\
57	12.5\\
58	12.75\\
59	13\\
60	13.25\\
61	13.5\\
62	13.75\\
63	14\\
64	14.25\\
65	14.5\\
66	14.75\\
67	15\\
68	15.25\\
69	15.5\\
70	15.75\\
71	16\\
72	16.25\\
73	16.5\\
74	16.75\\
75	17\\
76	17.25\\
77	17.5\\
78	17.75\\
79	18\\
80	18.25\\
81	18.5\\
82	18.75\\
83	19\\
84	19.25\\
85	19.5\\
86	19.75\\
87	20\\
88	20.25\\
89	20.5\\
90	20.75\\
91	21\\
92	21.25\\
93	21.5\\
94	21.75\\
95	22\\
96	22.25\\
97	22.5\\
98	22.75\\
99	23\\
100	23.25\\
};
\addlegendentry{Zero D.C.  Estim. Shifted by $\lambda=2$ }

\addplot [color=black, dashdotted, line width=0.3mm]
  table[row sep=crcr]{%
0	0.250000000000003\\
1	0.261904761904759\\
2	0.27488687782805\\
3	0.289084280506111\\
4	0.30465677330415\\
5	0.321789935316197\\
6	0.340699728632882\\
7	0.361637904282631\\
8	0.384898294416287\\
9	0.410824051836695\\
10	0.439815843160338\\
11	0.472340903362047\\
12	0.508942696573384\\
13	0.550250675123275\\
14	0.596989257904066\\
15	0.649984637087228\\
16	0.710167365998604\\
17	0.77856792316349\\
18	0.856301717178283\\
19	0.944539549201386\\
20	1.04445980863317\\
21	1.15718015139286\\
22	1.28366976379085\\
23	1.42464813975017\\
24	1.58048343341781\\
25	1.75110938871019\\
26	1.93597305845446\\
27	2.13408253318659\\
28	2.34389671062399\\
29	2.56399018767538\\
30	2.79335879425582\\
31	3.02357447335541\\
32	3.27463310512646\\
33	3.49129143357618\\
34	3.76203119847275\\
35	4.00709438150826\\
36	4.25406525801466\\
37	4.50226511117587\\
38	4.75122799715708\\
39	5.0006481823217\\
40	5.25033332930249\\
41	5.50016711088253\\
42	5.75008172602924\\
43	6.00003901212864\\
44	6.25001818740805\\
45	6.50000828536139\\
46	6.75000369019011\\
47	7.0000016076811\\
48	7.25000068544534\\
49	7.50000028613208\\
50	7.75000011699623\\
51	8.00000004687828\\
52	8.25000001841373\\
53	8.50000000709334\\
54	8.75000000268078\\
55	9.00000000099433\\
56	9.25000000036208\\
57	9.50000000012949\\
58	9.75000000004549\\
59	10.0000000000157\\
};
\addlegendentry{Conditional Expectation for $\lambda=5$}

\addplot [color=green,  line width=0.3mm]
  table[row sep=crcr]{%
0	0.224489795918367\\
1	0.244897959183673\\
2	0.26530612244898\\
3	0.285714285714286\\
4	0.306122448979592\\
5	0.326530612244898\\
6	0.346938775510204\\
7	0.36734693877551\\
8	0.387755102040816\\
9	0.408163265306122\\
10	0.428571428571429\\
11	0.448979591836735\\
12	0.469387755102041\\
13	0.489795918367347\\
14	0.510204081632653\\
15	0.530612244897959\\
16	0.551020408163265\\
17	0.571428571428571\\
18	0.591836734693878\\
19	0.612244897959184\\
20	0.63265306122449\\
21	0.653061224489796\\
22	0.673469387755102\\
23	0.693877551020408\\
24	0.714285714285714\\
25	0.73469387755102\\
26	0.755102040816326\\
27	0.775510204081633\\
28	0.795918367346939\\
29	0.816326530612245\\
30	0.836734693877551\\
31	0.857142857142857\\
32	0.877551020408163\\
33	0.897959183673469\\
34	0.918367346938776\\
35	0.938775510204082\\
36	0.959183673469388\\
37	0.979591836734694\\
38	1\\
39	1.02040816326531\\
40	1.04081632653061\\
41	1.06122448979592\\
42	1.08163265306122\\
43	1.10204081632653\\
44	1.12244897959184\\
45	1.14285714285714\\
46	1.16326530612245\\
47	1.18367346938776\\
48	1.20408163265306\\
49	1.22448979591837\\
50	1.24489795918367\\
51	1.26530612244898\\
52	1.28571428571429\\
53	1.30612244897959\\
54	1.3265306122449\\
55	1.3469387755102\\
56	1.36734693877551\\
57	1.38775510204082\\
58	1.40816326530612\\
59	1.42857142857143\\
60	1.44897959183673\\
61	1.46938775510204\\
62	1.48979591836735\\
63	1.51020408163265\\
64	1.53061224489796\\
65	1.55102040816327\\
66	1.57142857142857\\
67	1.59183673469388\\
68	1.61224489795918\\
69	1.63265306122449\\
70	1.6530612244898\\
71	1.6734693877551\\
72	1.69387755102041\\
73	1.71428571428571\\
74	1.73469387755102\\
75	1.75510204081633\\
76	1.77551020408163\\
77	1.79591836734694\\
78	1.81632653061224\\
79	1.83673469387755\\
80	1.85714285714286\\
81	1.87755102040816\\
82	1.89795918367347\\
83	1.91836734693878\\
84	1.93877551020408\\
85	1.95918367346939\\
86	1.97959183673469\\
87	2\\
88	2.02040816326531\\
89	2.04081632653061\\
90	2.06122448979592\\
91	2.08163265306122\\
92	2.10204081632653\\
93	2.12244897959184\\
94	2.14285714285714\\
95	2.16326530612245\\
96	2.18367346938776\\
97	2.20408163265306\\
98	2.22448979591837\\
99	2.24489795918367\\
100	2.26530612244898\\
};
\addlegendentry{Optimal Linear Estimator for $\lambda=5$}

\addplot [color=cyan, line width=0.3mm]
  table[row sep=crcr]{%
0	-4.75\\
1	-4.5\\
2	-4.25\\
3	-4\\
4	-3.75\\
5	-3.5\\
6	-3.25\\
7	-3\\
8	-2.75\\
9	-2.5\\
10	-2.25\\
11	-2\\
12	-1.75\\
13	-1.5\\
14	-1.25\\
15	-1\\
16	-0.75\\
17	-0.5\\
18	-0.25\\
19	0\\
20	0.25\\
21	0.5\\
22	0.75\\
23	1\\
24	1.25\\
25	1.5\\
26	1.75\\
27	2\\
28	2.25\\
29	2.5\\
30	2.75\\
31	3\\
32	3.25\\
33	3.5\\
34	3.75\\
35	4\\
36	4.25\\
37	4.5\\
38	4.75\\
39	5\\
40	5.25\\
41	5.5\\
42	5.75\\
43	6\\
44	6.25\\
45	6.5\\
46	6.75\\
47	7\\
48	7.25\\
49	7.5\\
50	7.75\\
51	8\\
52	8.25\\
53	8.5\\
54	8.75\\
55	9\\
56	9.25\\
57	9.5\\
58	9.75\\
59	10\\
60	10.25\\
61	10.5\\
62	10.75\\
63	11\\
64	11.25\\
65	11.5\\
66	11.75\\
67	12\\
68	12.25\\
69	12.5\\
70	12.75\\
71	13\\
72	13.25\\
73	13.5\\
74	13.75\\
75	14\\
76	14.25\\
77	14.5\\
78	14.75\\
79	15\\
80	15.25\\
81	15.5\\
82	15.75\\
83	16\\
84	16.25\\
85	16.5\\
86	16.75\\
87	17\\
88	17.25\\
89	17.5\\
90	17.75\\
91	18\\
92	18.25\\
93	18.5\\
94	18.75\\
95	19\\
96	19.25\\
97	19.5\\
98	19.75\\
99	20\\
100	20.25\\
};
\addlegendentry{Zero D.C. Estim. Shifted by $\lambda=5$}

\end{axis}

\end{tikzpicture}%
		\caption{Examples of conditional expectations  and linear estimators for $X$ distributed according to an exponential distribution with rate parameter $\alpha=3$ and $a=1$. } \label{fig:CEandLinear}
\end{figure}
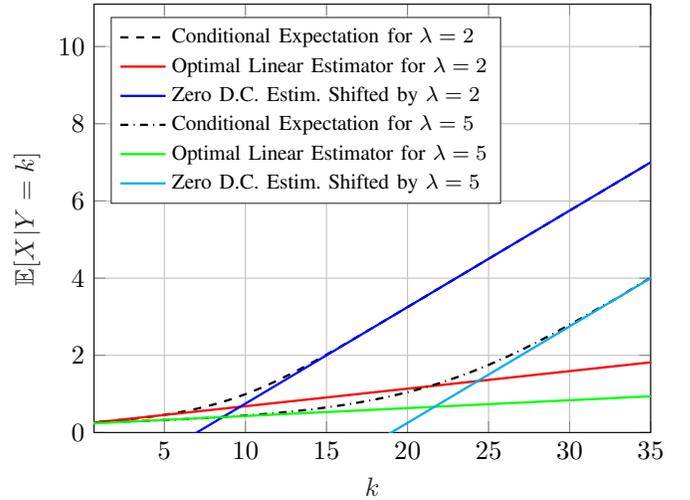

Fig.~\ref{fig:CEandLinear} compares the conditional expectation to the optimal linear estimator for an exponential random variable and shows that the conditional expectation can be approximated by a piece-wise linear function. More specifically, Fig.~\ref{fig:CEandLinear} shows that the optimal linear estimator is a good approximation of the conditional expectation for small values of count, and the optimal zero dark current linear estimator  shifted by the value of the dark current is a good approximation for large values of count.  
\begin{figure}[tb]  
	\centering   
%
%
\begin{tikzpicture}

\begin{axis}[%
width=7.4cm,
height=5.7cm,
at={(1.159in,0.77in)},
scale only axis,
xmin=0.656682027649771,
xmax=35,
xlabel style={font=\color{white!15!black}},
xlabel={$k$},
ymin=0,
ymax=11.1,
ylabel style={font=\color{white!15!black}},
ylabel={$\E[X|Y=k]$},
axis background/.style={fill=white},
xmajorgrids,
ymajorgrids,
legend pos=north west,
legend style={legend cell align=left, align=left, draw=white!15!black}
]
\addplot [color=black, solid,  line width=0.3mm]
  table[row sep=crcr]{%
0	0.25\\
1	0.5\\
2	0.75\\
3	1\\
4	1.25\\
5	1.5\\
6	1.75\\
7	2\\
8	2.25\\
9	2.5\\
10	2.75\\
11	3\\
12	3.25\\
13	3.5\\
14	3.75\\
15	4\\
16	4.25\\
17	4.5\\
18	4.75\\
19	5\\
20	5.25\\
21	5.5\\
22	5.75\\
23	6\\
24	6.25\\
25	6.5\\
26	6.75\\
27	7\\
28	7.25\\
29	7.5\\
30	7.75\\
31	8\\
32	8.25\\
33	8.5\\
34	8.75\\
35	9\\
36	9.25\\
37	9.5\\
38	9.75\\
39	10\\
40	10.25\\
41	10.5\\
42	10.75\\
43	11\\
44	11.25\\
45	11.5\\
46	11.75\\
47	12\\
48	12.25\\
49	12.5\\
50	12.75\\
51	13\\
52	13.25\\
53	13.5\\
54	13.75\\
55	14\\
56	14.25\\
57	14.5\\
58	14.75\\
59	15\\
};
\addlegendentry{$\lambda=0$}

\addplot [color=black,  dashed, line width=0.3mm]
  table[row sep=crcr]{%
0	0.25\\
1	0.277777777777777\\
2	0.310975609756097\\
3	0.350923482849605\\
4	0.399270482603816\\
5	0.458016606244889\\
6	0.529503770487892\\
7	0.616329399926177\\
8	0.721140522247445\\
9	0.846281655353462\\
10	0.993322128504657\\
11	1.16257664098021\\
12	1.35281277539743\\
13	1.56132926734424\\
14	1.78444178471602\\
15	2.01820177922015\\
16	2.25905964659576\\
17	2.50425438672973\\
18	2.75188902355175\\
19	3.00079420035606\\
20	3.25031847529539\\
21	3.5001374695329\\
22	3.74996857025491\\
23	4.00005617290663\\
24	4.25001957522359\\
25	4.50000653640201\\
26	4.75000209499854\\
27	5.00000064556972\\
28	5.25000019154265\\
29	5.50000005479636\\
30	5.75000001513423\\
31	6.00000000404028\\
32	6.25000000104374\\
33	6.50000000026119\\
34	6.75000000006338\\
35	7.00000000001492\\
36	7.25000000000342\\
37	7.50000000000076\\
38	7.75000000000016\\
39	8.00000000000004\\
40	8.25000000000001\\
41	8.5\\
42	8.75\\
43	9\\
44	9.25\\
45	9.5\\
46	9.75\\
47	10\\
48	10.25\\
49	10.5\\
50	10.75\\
51	11\\
52	11.25\\
53	11.5\\
54	11.75\\
55	12\\
56	12.25\\
57	12.5\\
58	12.75\\
59	13\\
};
\addlegendentry{$\lambda=2$}

\addplot [color=black, dashdotted, line width=0.3mm]
  table[row sep=crcr]{%
0	0.250000000000003\\
1	0.261904761904759\\
2	0.27488687782805\\
3	0.289084280506111\\
4	0.30465677330415\\
5	0.321789935316197\\
6	0.340699728632882\\
7	0.361637904282631\\
8	0.384898294416287\\
9	0.410824051836695\\
10	0.439815843160338\\
11	0.472340903362047\\
12	0.508942696573384\\
13	0.550250675123275\\
14	0.596989257904066\\
15	0.649984637087228\\
16	0.710167365998604\\
17	0.77856792316349\\
18	0.856301717178283\\
19	0.944539549201386\\
20	1.04445980863317\\
21	1.15718015139286\\
22	1.28366976379085\\
23	1.42464813975017\\
24	1.58048343341781\\
25	1.75110938871019\\
26	1.93597305845446\\
27	2.13408253318659\\
28	2.34389671062399\\
29	2.56399018767538\\
30	2.79335879425582\\
31	3.02357447335541\\
32	3.27463310512646\\
33	3.49129143357618\\
34	3.76203119847275\\
35	4.00709438150826\\
36	4.25406525801466\\
37	4.50226511117587\\
38	4.75122799715708\\
39	5.0006481823217\\
40	5.25033332930249\\
41	5.50016711088253\\
42	5.75008172602924\\
43	6.00003901212864\\
44	6.25001818740805\\
45	6.50000828536139\\
46	6.75000369019011\\
47	7.0000016076811\\
48	7.25000068544534\\
49	7.50000028613208\\
50	7.75000011699623\\
51	8.00000004687828\\
52	8.25000001841373\\
53	8.50000000709334\\
54	8.75000000268078\\
55	9.00000000099433\\
56	9.25000000036208\\
57	9.50000000012949\\
58	9.75000000004549\\
59	10.0000000000157\\
};
\addlegendentry{$\lambda=5$}

\end{axis}

\end{tikzpicture}%
		\caption{Examples of conditional expectations for $X$ distributed according to an exponential distribution with rate parameter $\alpha=3$ and $a=1$. } \label{fig:ConditionalExpectationExponential}
\end{figure}
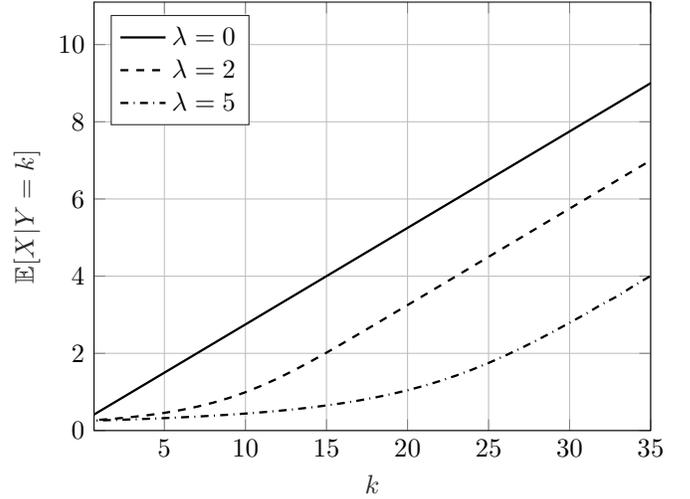%

\subsection{On the size of $\A$} 

Observe that according to Corollary~\ref{cor:InputXThm1}  $\A\X$ must have a product gamma distribution.    The following scenarios can be encountered:
\begin{itemize}
\item $\A$ is full rank.   In this case, the pdf  of $\X$ is  given by
\begin{align}
f_{\X}(\boldsymbol{x})= |{\rm  det}(\A)| f_{\U}( \A   \boldsymbol{x} ) 
\end{align}
where  $ f_{\U}( \cdot ) $ is the pdf of the product gamma distribution in  Corollary~\ref{cor:InputXThm1}. 
\item  $\A$ is a `fat' matrix (i.e.,  $k<n$).  In this case, there are several distributions on $\X$ that result in a product gamma distribution; and
\item $\A$ is a `thin' matrix (i.e.,  $k>n$). In this case, in general, it is not possible to generate a product distribution. 
\end{itemize}

\subsection{Comparison to the Gaussian Noise Case}
It is of some value to compare the result in the Poisson case to the Gaussian noise case.     The Gaussian counterpart of Theorem~\ref{thm:Main1} , which is a well-known result (see for example \cite[Lemma~5]{EPIallerton}), is given next.

\begin{theorem}\label{thm:LinearConditionGaussian} Suppose that $\A \in  \mathbb{R}^{k  \times n} $.  Let $\Y= \A \X+\Z$ where $\X \in \mathbb{R}^n$ and $\Z \sim \mathcal{N}(\boldsymbol{0}, \I)$ are independent. 
Then,
\begin{align}
\E[\X|\Y={\bf y}]=  \HH \boldsymbol{y}+\boldsymbol{c},  \forall  \boldsymbol{y} \in \mathbb{R}^n
\end{align} 
if and only if  $\X \sim \mathcal{N}( \boldsymbol{\mu},  \boldsymbol{ \mathsf{K}})$ such that
\begin{align}
\HH&= \boldsymbol{ \mathsf{K}} \A^T \left( \A  \boldsymbol{ \mathsf{K}} \A^T +\I \right)^{-1},\\
\boldsymbol{c}&=  \boldsymbol{\mu}- \HH  \A  \boldsymbol{\mu}. 
\end{align} 
In particular, $\A \X \sim \mathcal{N}( \A \mu,  \boldsymbol{\Sigma})  $ where  $ \boldsymbol{\Sigma}= (\I -\A \HH)^{-1} \A \HH$. 
\end{theorem} 

The key difference is that unlike in the Poisson noise case, in the Gaussian noise case the prior does not to have to be a product distribution.  In fact, in the Gaussian noise case, an arbitrary covariance matrix  on $\X$ results in a linear estimator.  Note that, while the distribution on $\A \X$ is unique, the distribution on $\X$ may not be unique and depends on the dimensionality of $\A$. 

To the best of our knowledge, for the Gaussian noise case  there exists only a scalar counterpart of Theorem~\ref{thm:QuantitativeRefinement}, which was shown in \cite[Lemma~4]{du2018strong}.  In order to make a proper comparison, the following result provides a vector Gaussian generalization.

\begin{theorem}\label{thm:GaussianStabilityResult}
Let  $\HH$ and $\boldsymbol{c}$ be as in Theorem~\ref{thm:LinearConditionGaussian}.   Denote by $ \phi_{\A \X}(\boldsymbol{t})$, $ \phi_{\Z}(\boldsymbol{t})$ and $\phi_{\Y}(\boldsymbol{t})$   the characteristic functions of $\A \X, \Z$ and $\Y$, respectively.   Assume that
\begin{align}
 \E \left[   \left \|\E[\X|\Y] -(\HH \Y+\boldsymbol{c}) \right \|^2 \right] \le \epsilon,
\end{align}  
for some $\epsilon \ge 0$. 
Then, for all $\boldsymbol{t} \in \mathbb{R}^k $
\begin{align}
 \frac{ \left|   \phi_{ \A \X}(\boldsymbol{t} )   -  \eu^{ -\frac{ \boldsymbol{t}^T \boldsymbol{\Sigma}  \boldsymbol{t} }{2}}   \right|  }{ \|\boldsymbol{t} \| } \le \frac{ \sqrt{\epsilon}  \|\A\|_{\star} }{  \sigma_{\min} \left ( \I-   \A \HH  \right)   \phi_{\Z} \left(   \boldsymbol{t} \right)} \label{eq:ControlOfCharInputGaussian}
 \end{align} 
 where  $ \boldsymbol{\Sigma}= ( \I-  \A  \HH )^{-1}   \A   \HH     $,  $\| \A\|_{\star}$ is the operator norm of $\A$, and  $ \sigma_{\min} \left ( \I-   \A \HH  \right)$ is the smallest singular value of $\I-   \A \HH$. 
 Consequently,
 \begin{align}
 \sup_{ \boldsymbol{t} \in \mathbb{R}^k } \frac{ \left|   \phi_{ \Y }(\boldsymbol{t} )   -  \eu^{ -\frac{ \boldsymbol{t}^T (\boldsymbol{\Sigma} + \I )  \boldsymbol{t} }{2}}   \right|  }{ \|\boldsymbol{t} \| } \le \frac{ \sqrt{\epsilon}  \|\A\|_{\star} }{  \sigma_{\min} \left ( \I-   \A \HH  \right)   }.   \label{eq:ControlOfCharOutputGaussian}
 \end{align} 
\end{theorem} 
\begin{IEEEproof}
See Appendix~\ref{app:thm:GaussianStabilityResult}. 
\end{IEEEproof} 

It is interesting to compare the Poisson result in \eqref{eq:ControllOfCharPoisson} to the Gaussian results in \eqref{eq:ControlOfCharInputGaussian} and \eqref{eq:ControlOfCharOutputGaussian}.   In particular, the Poisson result  appears to be stronger than the Gaussian result. In the Poisson case the control over the characteristic functions of the input in \eqref{eq:ControllOfCharPoisson} is uniform over all $\boldsymbol{t}$ (i.e., the domain of characteristic functions), but in the  Gaussian counterpart in \eqref{eq:ControlOfCharInputGaussian} such a bound is not uniform over all $\boldsymbol{t}$. In the Gaussian case, we do get a  uniform bound, but only for  the characteristic functions of the output  as shown in \eqref{eq:ControlOfCharOutputGaussian}.

\begin{appendices} 
\section{Proof of Lemma~\ref{lem:ConditionalCovMatrix}} 
\label{app:lem:ConditionalCovMatrix}

First, compute the cross-correlation term 
\begin{align}
\E[U_i U_j|\Y= \boldsymbol{y} ] 
&= \frac{\E[ U_i  U_j P_{\Y|\U}(\boldsymbol{y} |\U)]}{ P_\Y(\boldsymbol{y}) }\\
&=\frac{  (y_i+1) (y_j+1) P_\Y(\boldsymbol{y} +  \boldsymbol{1}_{i} + \boldsymbol{1}_{j})}{ P_\Y(\boldsymbol{y}) }. 
\end{align}
Therefore, by using Lemma~\ref{lem:EmpericalBayes} 
\begin{align}
&[ \mathsf{\boldsymbol{Var}}(\U|\Y=\boldsymbol{y}) ]_{ij} \notag\\
& =  \frac{  (y_i+1) (y_j+1) P_\Y(\boldsymbol{y} +  \boldsymbol{1}_{i} + \boldsymbol{1}_{j})}{ P_\Y(\boldsymbol{y}) }  \notag\\
& -  \frac{  (y_i+1) (y_j+1) P_\Y(\boldsymbol{y} +  \boldsymbol{1}_{i}) P_\Y(\boldsymbol{y} + \boldsymbol{1}_{j})}{ P_\Y(\boldsymbol{y})  P_\Y(\boldsymbol{y}) } \\
&=  \hspace{-0.05cm}\E[ U_i | \Y= \boldsymbol{y} ]   \left(\hspace{-0.05cm}   \frac{  (y_j+1) P_\Y(\boldsymbol{y} +  \boldsymbol{1}_{i} + \boldsymbol{1}_{j})}{  P_\Y(\boldsymbol{y} +\boldsymbol{1}_{i}) }   \hspace{-0.05cm} -  \E[ U_j | \Y= \boldsymbol{y} ]  \hspace{-0.05cm}  \right) \\
&= \E[ U_i | \Y= \boldsymbol{y} ]   \left(  \E[ U_j | \Y= \boldsymbol{y}+\boldsymbol{1}_{i} ]    -  \E[ U_j | \Y= \boldsymbol{y} ]   \right). 
\end{align} 
This concludes the proof.

\section{Proof of Theorem~\ref{thm:GaussianStabilityResult}}
\label{app:thm:GaussianStabilityResult}
The proof for the Gaussian case is very similar to the Poisson case.  We start with the following lemma. 

 \begin{lem} Let $\boldsymbol{\Sigma}$ be some covariance matrix and $\phi_{\X} \left( \boldsymbol{t} \right)$ be the characteristic function of random vector $\X \in \mathbb{R}^n$.  Then,  for every $  \boldsymbol{t}\in \mathbb{R}^n$ 
 \begin{align}
  \left| \phi_{\X} \left( \boldsymbol{t} \right)  - \eu^{ -\frac{ \boldsymbol{t}^T\boldsymbol{\Sigma}  \boldsymbol{t} }{2}} \right| \le   \|\boldsymbol{t} \|   \max_{\tau \in [0,1]}  \|  \nabla \phi_{\X} \left(  \tau \boldsymbol{t}  \right)   +  \boldsymbol{\Sigma}  \tau \boldsymbol{t} \phi_{\X} \left(  \tau \boldsymbol{t}  \right)      \|.  \label{eq:DerivativeBoundOnDiffChar}
 \end{align} 
 \end{lem} 
 \begin{IEEEproof}
 Let $\boldsymbol{r}(\tau)= \tau \boldsymbol{t} $ 
 \begin{align}
  & \left| \phi_{\X} \left( \boldsymbol{t} \right) \eu^{ \frac{ \boldsymbol{t}^T \boldsymbol{\Sigma}  \boldsymbol{t} }{2}}   -  1 \right|  \\
 &=    \left|  \int_0^1   \nabla \phi_{\X} \left(\boldsymbol{r}(\tau) \right) \eu^{ \frac{\boldsymbol{r}(\tau)^T \boldsymbol{\Sigma} \boldsymbol{r}(\tau)  }{2}}  \boldsymbol{\cdot} \dot{\boldsymbol{r}}(\tau)  {\rm d} \tau  \right|  \label{eq:FTCgaussianApplication} \\
 &\le     \int_0^1    \left|  \nabla \phi_{\X} \left(\boldsymbol{r}(\tau) \right) \eu^{ \frac{\boldsymbol{r}(\tau)^T \boldsymbol{\Sigma} \boldsymbol{r}(\tau)  }{2}}  \boldsymbol{\cdot} \dot{\boldsymbol{r}}(\tau)  \right|   {\rm d} \tau \label{eq:MOdulusInequalityAppplication}  \\
    & \le    \|\boldsymbol{t} \|      \int_0^1   \eu^{  \tau^2 \frac{ \boldsymbol{t}^T \boldsymbol{\Sigma}  \boldsymbol{t}  }{2}}  \|  \nabla \phi_{\X} \left(  \tau \boldsymbol{t}  \right)   +   \boldsymbol{\Sigma}  \tau \boldsymbol{t} \phi_{\X} \left(  \tau \boldsymbol{t}  \right)      \|  {\rm d} \tau  \label{eq:NormBoundsCauchySchwarz}\\
    & \le   \|\boldsymbol{t} \|   \max_{\tau \in [0,1]}  \|  \nabla \phi_{\X} \left(  \tau \boldsymbol{t}  \right)   +   \tau   \boldsymbol{\Sigma} \boldsymbol{t} \phi_{\X} \left(  \tau \boldsymbol{t}  \right)      \| 
       \int_0^1   \eu^{  \tau^2 \frac{ \boldsymbol{t}^T \boldsymbol{\Sigma}  \boldsymbol{t}  }{2}}    {\rm d} \tau \\
          & \le   \|\boldsymbol{t} \|   \max_{\tau \in [0,1]}  \|  \nabla \phi_{\X} \left(  \tau \boldsymbol{t}  \right)   +   \tau \boldsymbol{\Sigma}   \boldsymbol{t} \phi_{\X} \left(  \tau \boldsymbol{t}  \right)      \| 
       \eu^{   \frac{ \boldsymbol{t}^T \boldsymbol{\Sigma}  \boldsymbol{t}  }{2}}  ,   \label{eq:BoundOnRationMinus1}
     \end{align} 
     where  \eqref{eq:FTCgaussianApplication} follows from the fundamental theorem of calculus for line integrals;  \eqref{eq:MOdulusInequalityAppplication} follows from modulus inequality;  and \eqref{eq:NormBoundsCauchySchwarz}  is a consequence of using  $\dot{\boldsymbol{r}}(\tau)=  \boldsymbol{t} $, $\nabla   \eu^{  \frac{  \boldsymbol{t}^T \boldsymbol{\Sigma}  \boldsymbol{t}   }{2}} = \boldsymbol{\Sigma}  \boldsymbol{t}   \eu^{  \frac{  \boldsymbol{t}^T \boldsymbol{\Sigma}  \boldsymbol{t}   }{2}}$ and the Cauchy-Schwarz inequality to produces the following sequence of bounds: 
     \begin{align}
    & \left|  \nabla \phi_{\X} \left(\boldsymbol{r}(\tau) \right) \eu^{ \frac{\boldsymbol{r}(\tau)^T \boldsymbol{\Sigma} \boldsymbol{r}(\tau)  }{2}}  \boldsymbol{\cdot}\dot{\boldsymbol{r}}(\tau)  \right|   \notag\\
     &=     \eu^{  \frac{ \tau^2 \boldsymbol{t}^T \boldsymbol{\Sigma}  \boldsymbol{t}   }{2}}  \left| \left(  \nabla \phi_{\X} \left( \tau \boldsymbol{t}  \right)   +  \tau \boldsymbol{\Sigma}  \boldsymbol{t} \phi_{\X} \left( \tau \boldsymbol{t}  \right)    \right) \boldsymbol{\cdot} \boldsymbol{t}     \right| \\
     & \le   \eu^{  \frac{ \tau^2 \boldsymbol{t}^T \boldsymbol{\Sigma}  \boldsymbol{t}   }{2}}  \left \|  \nabla \phi_{\X} \left( \tau \boldsymbol{t}  \right)   +  \tau  \boldsymbol{\Sigma}  \boldsymbol{t}    \phi_{\X} \left( \tau \boldsymbol{t}  \right)   \right \|  \|  \boldsymbol{t}  \|. 
     \end{align} 
This concludes the proof. 
     \end{IEEEproof}

Now,  using  the orthogonality principle  observe that 
\begin{align}
\boldsymbol{0}&=\E \left[ ( \A \X-\E[ \A \X| \Y]) \eu^{ i \boldsymbol{t}^T \Y } \right]\\
&= \E \left[ ( \A\X-   \A \HH \Y + \A \HH \Y-\E[ \A\X| \Y]) \eu^{ i \boldsymbol{t}^T \Y } \right]\\
&= \E \left[ ( \A \X-   \A \HH \Y)\eu^{ i \boldsymbol{t}^T \Y } \right ]  \notag\\
&+ \E \left[  ( \A \HH \Y-\E[ \A \X| \Y]) \eu^{ i \boldsymbol{t}^T \Y } \right]. \label{eq:EquaivalentBetweenLinearAndNonLinear}
\end{align} 

Moreover, the first term in \eqref{eq:EquaivalentBetweenLinearAndNonLinear} can be computed in terms of characteristic functions as follows:
\begin{align}
&\E \left[ ( \A \X-  \A \HH\Y)\eu^{ i \boldsymbol{t}^T \Y } \right ]  \notag \\
&= \E \left[ ( \A \X-   \A\HH \A \X-  \A \HH \Z)\eu^{ i \boldsymbol{t}^T \Y } \right ] \\
& =\E \left[  \left( \I-  \A \HH  \right)  \A \X \eu^{ i \boldsymbol{t}^T \Y} - \A \HH \Z\eu^{ i \boldsymbol{t}^T \Y } \right ] \\
&= \E \left[ \left ( \I-  \A \HH \right)  \A \X  \eu^{ i \boldsymbol{t}^T \A \X}  \right]  \E \left[ \eu^{ i \boldsymbol{t}^T \Z} \right] \\
& - \E \left[ \A \HH \Z \eu^{ i \boldsymbol{t}^T \Z} \right ] \E \left[\eu^{ i \boldsymbol{t}^T \A \X } \right]  \label{eq:UsingIndependence} \\
&=    ( \I-   \A \HH)  \E \left [ \A  \X \eu^{ i \boldsymbol{t}^T \A \X} \right] \phi_{\Z}(\boldsymbol{t})  \notag\\
&-   \A \HH \E \left[  \Z \eu^{ i \boldsymbol{t}^T \Z} \right]    \phi_{ \A \X} \left( \boldsymbol{t} \right)  \\
&=  ( \I-  \A \HH ) \frac{1}{i}  \nabla \phi_{ \A \X}(   \boldsymbol{t})   \phi_{\Z}(\boldsymbol{t}) - \A  \HH  \frac{1}{i}  \nabla \phi_{\Z}(\boldsymbol{t})    \phi_{ \A \X} \left(  \boldsymbol{t} \right) \label{eq:GradientResults}  \\
& =  ( \I-  \A \HH) (-i)  \nabla\phi_{ \A \X}(  \boldsymbol{t})   \phi_{\Z}(\boldsymbol{t}) +  \A \HH \boldsymbol{t} (-i)   \phi_{\Z}(\boldsymbol{t})    \phi_{ \A \X} \left( \boldsymbol{t} \right)   \label{eq:GradientOfPhiZ}\\
&=   (-i) \left( ( \I-   \A \HH )   \nabla\phi_{ \A \X}(   \boldsymbol{t})    + \A  \HH \boldsymbol{t}       \phi_{ \A \X} \left( \boldsymbol{t} \right)  \right)    \phi_{\Z}(\boldsymbol{t}),\label{eq:ExpressionForTheLinearDiff}
\end{align}
where \eqref{eq:UsingIndependence} follows from the independence of $\X$ and $\Z$; \eqref{eq:GradientResults} follows by observing that  $\nabla \phi_{ \A \X}( \boldsymbol{t})=  \E[  i \A \X \eu^{ i \boldsymbol{t}^T  \A \X} ]$ and  $\nabla \phi_{\Z}( \boldsymbol{t})= \E[  i \Z \eu^{ i \boldsymbol{t}^T  \Z} ]$; and \eqref{eq:GradientOfPhiZ} follows by using that  $\nabla\phi_{\Z}( \boldsymbol{t})= - \boldsymbol{t}  \phi_{\Z}( \boldsymbol{t}) $.

Next, by using \eqref{eq:EquaivalentBetweenLinearAndNonLinear} and \eqref{eq:ExpressionForTheLinearDiff}, and applying the norm on both sides we get that 
\begin{align}
&\left \|   \E \left[ \A ( \HH \Y-\E[\X| \Y])^T \eu^{ i \boldsymbol{t}^T \Y } \right] \right\| \notag\\
&= \left\|   \phi_{\Z}(\boldsymbol{t}) \left( ( \I-   \A \HH )  \nabla \phi_{ \A \X}(  \boldsymbol{t})   +       \A  \HH  \boldsymbol{t} \phi_{ \A \X} \left(  \boldsymbol{t} \right)  \right)  \right\|\\
&=   \phi_{\Z}(\boldsymbol{t})  \left\|  ( \I-  \A \HH )  \nabla \phi_{ \A \X}(  \boldsymbol{t})   +       \A   \HH  \boldsymbol{t} \phi_{ \A \X} \left( \boldsymbol{t} \right)   \right\|. \label{eq:AfterTheNorm}
\end{align}
Furthermore, by using the Cauchy-Schwarz  inequality in \eqref{eq:AfterTheNorm}
\begin{align}
& \sqrt{\E \left[ \left \| \A ( \HH \Y-\E[\X| \Y])  \right\|^2 \right]} \notag\\
 & \ge  \phi_{\Z}(\boldsymbol{t})  \left\|  ( \I-  \A \HH )  \nabla \phi_{ \A \X}(  \boldsymbol{t})   +       \A  \HH  \boldsymbol{t} \phi_{ \A \X} \left( \boldsymbol{t} \right)   \right\|,\\
& \ge    \phi_{\Z}(\boldsymbol{t}) \sigma_{\min}( \I-  \A \HH ) \notag\\
& \quad \cdot   \left\|    \nabla \phi_{ \A \X}(  \boldsymbol{t})   +      ( \I-  \A \HH )^{-1}   \A  \HH  \boldsymbol{t} \phi_{ \A \X} \left( \boldsymbol{t} \right)   \right\|, \label{eq:SpectralBound}  
\end{align} 
where  \eqref{eq:SpectralBound} follows by using the fact that $(\I-  \A \HH )$ is invertible and the inequality $\| \A \boldsymbol{x} \|  \ge   \sigma_{\min}(\A) \| \boldsymbol{x} \|, \forall \boldsymbol{x}$ where $ \sigma_{\min}(\A)$ is the small singular value of $\A$. 

Combining bounds in \eqref{eq:DerivativeBoundOnDiffChar} and \eqref{eq:SpectralBound} and using the bound $\| \A \boldsymbol{x} \|  \le   \|\A\|_{\star} \| \boldsymbol{x} \|, \forall \boldsymbol{x}$ we have that
 \begin{align}
\frac{  \left| \phi_{ \A \X} \left( \boldsymbol{t} \right)  - \eu^{ -\frac{ \boldsymbol{t}^T \boldsymbol{\Sigma}  \boldsymbol{t} }{2}} \right| }{  \|\boldsymbol{t} \|}   &   \le   \frac{ \sqrt{\E \left[ \left \| \A ( \HH \Y-\E[\X| \Y])  \right\|^2 \right]} }{  \sigma_{\min} \left ( \I- \A \HH  \right)   \phi_{\Z} \left(  \boldsymbol{t} \right)}\\
 & \le   \frac{  \|\A\|_{\star}  \sqrt{\E \left[ \left \|  \HH \Y-\E[\X| \Y] \right\|^2 \right]} }{  \sigma_{\min} \left ( \I- \A \HH  \right)   \phi_{\Z} \left(  \boldsymbol{t} \right)},
 \end{align} 
 where  $ \boldsymbol{\Sigma}= ( \I-  \A \HH )^{-1}     \A  \HH $. This concludes the proof.

\end{appendices}

\bibliography{refs}
\bibliographystyle{IEEEtran}
\end{document}